\documentclass[aps,prd,onecolumn,longbibliography,titlepage,nofootinbib]{revtex4-1}

\usepackage{amsthm}
\usepackage{amssymb}   
\usepackage{dsfont}
\usepackage{mathtools} 
\usepackage{hyperref}
\hypersetup{colorlinks=true,linkcolor=blue,urlcolor=blue,citecolor=blue}
\usepackage{accents}
\usepackage{tensor}
\usepackage{xcolor}
\usepackage{graphicx}
\usepackage[cal=boondox]{mathalfa}
\usepackage{lipsum}
\usepackage{relsize}
\usepackage{soul}
\usepackage{comment}
\usepackage{nameref}
\usepackage{lineno}
\usepackage{orcidlink}

\begin{document}
	
\title{Combining Symmetries and Helmholtz's Conditions to Construct Lagrangians}

\author{Merced Montesinos$^1$ \orcidlink{0000-0002-4936-9170}}
\email{merced.montesinos@cinvestav.mx}

\author{Diego Gonzalez$^{1,2}$ \orcidlink{0000-0002-0206-7378}}
\email{dgonzalezv@ipn.mx}

\author{Jorge Meza$^1$ \orcidlink{0009-0002-5022-219X}}
\email{jorge.meza@cinvestav.mx}

\affiliation{$^1$Department of Physics, Center for Research and Advanced Studies, 2508 National Polytechnic Institute Avenue, San Pedro Zacatenco, 07360, Gustavo A. Madero, Mexico City, Mexico}

\affiliation{$^2$Higher School of Mechanical and Electrical Engineering, National Polytechnic Institute, Adolfo L\'opez Mateos Professional Unit, 07738, Gustavo A. Madero, Mexico City, Mexico}

\date{\today}
	
\begin{abstract}
We present new relations derived from Noether’s identity that reveal the compatibility between the components of the Hessian matrix of the Lagrangian, the infinitesimal symmetry transformation of the configuration variables and time, and a constant of motion. Using these relations, we develop two new methods to incorporate symmetry requirements directly into the inverse problem of mechanics, thereby restricting the set of acceptable Lagrangians. We accomplish this by combining these relations with Helmholtz’s conditions, which allow us to construct Lagrangians whose actions exhibit specific symmetries from the outset. The theory is illustrated with one- and two-dimensional examples.
\end{abstract}
	
\maketitle

\section{Introduction}
Dynamical systems remain a very active field of research in both physics and mathematics. For instance, non-Hamiltonian methods for counting physical degrees of freedom were reported recently in~\cite{Mont1JMP,Mont2JMP}. In this article, we focus on the study of symmetries and the inverse problem of mechanics~\cite{Bateman,Douglas_paper}, an issue with deep implications for the construction of physically meaningful actions~\cite{Marmo,Helliwell_Sahakian_book}.  

Suppose that we are given a system of equations of motion and we are interested in finding a Lagrangian whose corresponding Euler–Lagrange equations reproduce them. We can proceed by inspection, as usual, to find such a Lagrangian and then determine the symmetries of the resulting action constructed from it. Alternatively, we can employ a fully systematic approach that uses Helmholtz's conditions as a starting point to determine the desired Lagrangian, and thus the corresponding action. We remind the reader that Helmholtz's conditions are necessary and sufficient for the existence of a Lagrangian: necessity was shown by Helmholtz himself~\cite{Helmholtz1887}, and sufficiency was proved by Douglas~\cite{Douglas_paper} (inverse problem of mechanics). Therefore, if Helmholtz's conditions are not satisfied, then there is no Lagrangian leading to the equations of motion by means of the Euler-Lagrange equations. However, if Helmholtz's conditions are satisfied, then there might be more than one solution. It is important to recall that each of these solutions leads to a distinct Lagrangian, that is, the Lagrangians do not differ from each other by a total derivative with respect to the time variable $t$ of a function $f(q^i,t)$ depending only on coordinates $q^i$ and time $t$. Up to this point, we have described the inverse problem of mechanics, namely, the problem of finding a Lagrangian from a given system of equations of motion. Once the Lagrangian has been obtained, the next natural step is to study the symmetries of the corresponding action~\cite{Noether,EmmyNoether,Bessel-Hagen}.

In this article, we propose two different approaches to the inverse problem of mechanics. In the first approach, we combine symmetries with the inverse problem of mechanics from the very beginning. From scratch, we can only talk about the symmetries of the equations of motion as a starting point if the only thing we have at hand is precisely the equations of motion~\cite{Marmo}. The goal is then to look for a Lagrangian that incorporates some of the symmetries of the equations of motion. The reason for doing this is that, in physics, we are usually not interested in finding all Lagrangians that merely lead to a given system of equations of motion, but rather one whose action exhibits a specific symmetry~\cite{Helliwell_Sahakian_book}. Additionally, since in physics we are usually interested in finding a Lagrangian (and hence an action) endowed with a specific symmetry that provides a desired constant of motion through the application of Noether's first theorem, the second approach goes a step further by embedding not only the symmetry but also the associated constant of motion directly into the inverse problem. In this article, we address these two problems by answering the following question: How must Helmholtz's conditions~\cite{Helmholtz1887,Douglas_paper} be modified for the action constructed with the resulting Lagrangian to exhibit the desired symmetry (or symmetries) from the outset?  

To accomplish the goals outlined in the previous paragraph, we first derive in Section~\ref{results} certain relations that arise from Noether's identity for rigid symmetries~\cite{Noether,EmmyNoether,Bessel-Hagen} (i.e., gauge transformations are not considered in this article). A relevant feature of these relations is that they involve the components of the Hessian matrix associated with a Lagrangian, a fact that is crucial to achieve the aforementioned goals, as this matrix lies at the core of Helmholtz’s conditions~\cite{Helmholtz1887,Douglas_paper}. 

We hope that this article fills a gap in the literature on the subject. 

%%%%%%%%%%%%%%%
\section{Noether's identity implies new relations}\label{results}

We begin by recalling Noether's first theorem in order to establish the notation and conventions used throughout this article. In particular, we use Einstein's summation convention in expressions that have repeated indices.

{\bf Noether's first theorem}. If the `off-shell' variation of the action equals~\cite{Noether,EmmyNoether,Bessel-Hagen}:
\begin{eqnarray} \label{Noether_hypothesis}
\Delta S & = & \Delta \int^{t_2}_{t_1}
{\cal L}(q, {\dot q},t) dt =  \int^{t_2}_{t_1} \frac{d {\mathcal F}}{dt} dt,
\end{eqnarray}
with ${\mathcal F} = {\mathcal F}_a (q(t),t) \epsilon^a$, under the `small' infinitesimal transformation\footnote{Some authors consider a version of Noether's first theorem in which $\Delta t$ and $\Delta q^i$ are also allowed to depend on the velocities ${\dot q}^i$. In such a case, also ${\mathcal F} = {\mathcal F}_a (q(t), {\dot q} (t), t) \epsilon^a$ must hold (see, for instance, Refs.~\cite{olver_book,Harvey}). In this article, we restrict ourselves to the standard case in physics, in which $\Delta t$, $\Delta q^i$, and ${\mathcal F}$ are independent of ${\dot q}^i$ (see, for instance,~\cite{Helliwell_Sahakian_book,Gerardo_book}). See, however, Section~\ref{CR}.}
\begin{align}
t' & =  t +\Delta t, \quad \Delta t = T_a (q(t),t) \epsilon^a, \quad  a=1, \ldots, r, \label{timetransf} \\
{q'}^i (t') & =  q^i (t) + \Delta q^i, \quad \Delta q^i = Q^i_a (q(t),t) \epsilon^a, \quad i=1, \ldots, n,  \label{spacetransf}
\end{align}
then
\begin{equation}\label{NC1}
\left ( \frac{\partial {\mathcal L}}{\partial q^i} - \frac{d}{dt} \frac{\partial {\mathcal L}}{\partial {\dot q}^i}  \right )  \delta q^i 
+ \frac{d}{d t}  \left ( \frac{\partial {\mathcal L}}{\partial {\dot q}^i} \delta q^i + {\mathcal L} \Delta t - {\mathcal F}  \right ) =0.
\end{equation}
The infinitesimal transformation is `small' in the sense that if the infinitesimal constant parameters $\epsilon^a$ that appear linearly on the right-hand side of~\eqref{timetransf} and~\eqref{spacetransf} are set to zero, i.e., $\epsilon^a=0$, we get the identity transformation.  Using the expression $\Delta= \delta + \Delta t \frac{d}{dt}$, which relates the total variation $\Delta$ and the variation $\delta$, which fulfills $\delta \frac{d}{dt} - \frac{d}{dt} \delta=0$, the expression~\eqref{NC1} can also be written as
\begin{eqnarray}\label{NC2}
&&\left ( \frac{\partial {\mathcal L}}{\partial q^i} - \frac{d}{dt} \frac{\partial {\mathcal L}}{\partial {\dot q}^i} \right ) \delta q^i + \frac{d}{d t}  \bigg[ \frac{\partial {\mathcal L}}{\partial {\dot q}^i} \Delta q^i + \left (  {\mathcal L} - \frac{\partial {\mathcal L}}{\partial {\dot q}^i} {\dot q}^i \right )  \Delta t - {\mathcal F}  \bigg] =0.
\end{eqnarray}

The relation~\eqref{NC1} or~\eqref{NC2} is referred to as {\it Noether's identity}. Note that it is an `off-shell' relation, i.e., it holds without invoking the equations of motion. From~\eqref{NC1}, it is clear that 
\begin{eqnarray}\label{CM1}
 \frac{\partial {\mathcal L}}{\partial {\dot q}^i} \delta q^i + {\mathcal L} \Delta t - {\mathcal F}
\end{eqnarray}
is a constant of motion, which according to~\eqref{NC2} can also be written as 
\begin{eqnarray}\label{CM2}
 \frac{\partial {\mathcal L}}{\partial {\dot q}^i} \Delta q^i + \left (  {\mathcal L} - \frac{\partial {\mathcal L}}{\partial {\dot q}^i} {\dot q}^i \right )  \Delta t - {\mathcal F}.
\end{eqnarray}

Before addressing the subject of symmetries and the inverse problem of mechanics, we first derive certain new and relevant relations that arise from Noether's identity~\eqref{NC1} or~\eqref{NC2}. The new relations reported below are not included in Noether's seminal article on symmetries~\cite{Noether,EmmyNoether,Bessel-Hagen} and, of course, have not reported in the vast literaure on the subject before this article.

{\bf First relations}. Let us denote the constant of motion~\eqref{CM2} involved in Noether's identity~\eqref{NC2} as follows:
\begin{eqnarray}\label{def_C}
{\mathcal C} :=  \frac{\partial {\mathcal L}}{\partial {\dot q}^i} \Delta q^i + \left (  {\mathcal L} - \frac{\partial {\mathcal L}}{\partial {\dot q}^i} {\dot q}^i \right )  \Delta t - {\mathcal F}.
\end{eqnarray}

Taking into account that $\Delta q^i$, $\Delta t$, and ${\mathcal F}$ are independent of ${\dot q}^i$, from~\eqref{def_C} we get
\begin{eqnarray}\label{previous_delta_q}
\frac{\partial {\mathcal C}}{\partial {\dot q}^i} = M_{ij} \delta q^j, 
\end{eqnarray}
with $M_{ij} =\frac{\partial^2 {\mathcal L}}{\partial {\dot q}^i \partial {\dot q}^j}$. Equivalently, 
\begin{eqnarray}\label{delta_q}
\delta q^i &=& M^{ij} \frac{\partial {\mathcal C}}{\partial {\dot q}^j},
\end{eqnarray}
where $M^{ij}$ is the inverse of $M_{ij}$. This is a remarkable expression that encodes the variation of the configuration variables in terms of the geometrical structure $M^{ij}$ and the constant of motion ${\mathcal C}$.

In general, the left-hand side of~\eqref{delta_q} is also equal to $\delta q^i= \Delta q^i - {\dot q}^i \Delta t$. Equating both expressions and calculating the partial derivative with respect to the velocities ${\dot q}^i$, under the assumption that $\Delta q^i$ and $\Delta t$ are independent of ${\dot q}^i$, we derive
\begin{eqnarray}\label{relevant1}
\Delta t &=& - \frac{1}{n} \frac{\partial}{\partial {\dot q}^i} \left ( M^{ij} \frac{\partial {\mathcal C}}{\partial {\dot q}^j} \right ).
\end{eqnarray}
Therefore, using~\eqref{delta_q} and~\eqref{relevant1}, we obtain
\begin{eqnarray}\label{relevant2}
\Delta q^i = \delta q^i + {\dot q}^i \Delta t  = M^{ij} \frac{\partial {\mathcal C}}{\partial {\dot q}^j} - \frac{1}{n} {\dot q}^i \frac{\partial}{\partial {\dot q}^j} \left ( M^{jk} \frac{\partial {\mathcal C}}{\partial {\dot q}^k} \right ).
\end{eqnarray}
Note that expressions~\eqref{relevant1} and~\eqref{relevant2} beautifully combine three key elements: the infinitesimal symmetry transformation $\Delta t$ and $\Delta q^i$, the Lagrangian ${\mathcal L}$ (contained in $M^{ij}$), and the constant of motion ${\mathcal C}$. In other words, the infinitesimal symmetry transformation $\Delta t$ and $\Delta q^i$ is expressed in terms of the constant of motion ${\mathcal C}$ and $M_{ij}$. 

Considering that $\Delta t = T_a  (q,t) \epsilon^a$, $\Delta q^i = Q^i_a (q,t) \epsilon^a$, ${\mathcal F}= {\mathcal F}_a (q,t) \epsilon^a$, and ${\mathcal C}={\mathcal C}_a (q, {\dot q}, t) \epsilon^a$, where $\epsilon^a$ are the infinitesimal constant parameters corresponding to each infinitesimal transformation, expressions~\eqref{relevant1} and~\eqref{relevant2} can also be written as follows:
\begin{eqnarray}
T_a &=& - \frac{1}{n} \frac{\partial}{\partial {\dot q}^i} \left ( M^{ij} \frac{\partial {\mathcal C}_a}{\partial {\dot q}^j} \right ), \label{relevant3} \\
Q^i_a &=& M^{ij} \frac{\partial {\mathcal C}_a}{\partial {\dot q}^j} - \frac{1}{n} {\dot q}^i \frac{\partial}{\partial {\dot q}^j} \left ( M^{jk} \frac{\partial {\mathcal C}_a}{\partial {\dot q}^k} \right ). \label{relevant4}
\end{eqnarray}

We make three remarks as follows:

First, note that expressions~\eqref{relevant1} and~\eqref{relevant2}, or their equivalents~\eqref{relevant3} and~\eqref{relevant4}, clearly display the fact that two Lagrangians leading to the same equations of motion, but corresponding to distinct $M_{ij}$, can have the same symmetry $\Delta q^i$ and $\Delta t$, provided that the associated constants of motion also differ. Let us illustrate this point:

The Lagrangians ${\mathcal L} =\frac12 m {\dot x}^2 e^{\lambda t}$ and ${\mathcal L} = m({\dot x} \ln{|\dot x|} - \lambda x)$, where $m$ is the mass and $\lambda$ is a positive constant, both give the equation of motion ${\ddot x} + \lambda {\dot x}=0$ of a damped ``free'' particle. Note that $M_{11}= m e^{\lambda t}$ for the first, while $M_{11}=m/{\dot x}$ for the second Lagrangian. The infinitesimal transformation $\Delta t = - e^{\lambda t} \epsilon$ and $\Delta x =0$ is a strict symmetry (${\mathcal F}=0$) of the action constructed with the first Lagrangian, while the same infinitesimal transformation is also a symmetry of the action constructed with the second Lagrangian, but gives rise to a boundary term (${\mathcal F}=\lambda m x e^{\lambda t}\epsilon $). For the first Lagrangian, the constant of motion is ${\mathcal C} = \frac12 m {\dot x}^2 e^{2 \lambda t} \epsilon$ (confront with Ref.~\cite{Havas}), while for the second Lagrangian the constant of motion is ${\mathcal C}= m{\dot x} e^{\lambda t} \epsilon$. Thus, the same symmetry is associated with two different constants of motion arising from different Lagrangians.\footnote{Note incidentally, that in page 153 of Ref.~\cite{Havas} it is stated that the constant of motion ${\mathcal C} = \frac12 m {\dot x}^2 e^{2 \lambda t} \epsilon$ does not follow from the application of Noether's first theorem to the action constructed with the Lagrangian ${\mathcal L} =\frac12 m {\dot x}^2 e^{\lambda t}$, which is incorrect, as we have shown. In fact, the finite version of the infinitesimal transformation $\Delta t = - e^{\lambda t} \epsilon$ and $\Delta x =0$ is
\begin{eqnarray}\label{finite_t}
t'= - \frac{1}{\lambda} \ln{\left ( e^{-\lambda t} + \alpha \lambda \right )}, \quad x' (t')= x(t), 
\end{eqnarray}
where $\alpha$ is an arbitrary constant parameter (provided that the previous expression makes sense).  Note that if $\alpha=0$, then $t'=t$, so the transformation~\eqref{finite_t} has the desired property. It has inverse and the composition of two transformations closes.}

Second, note also that the expressions~\eqref{relevant1} and~\eqref{relevant2}, or their equivalents~\eqref{relevant3} and~\eqref{relevant4}, clearly show that two distinct symmetries, each of them corresponding to two different Lagrangians, can be associated with the same constant of motion ${\mathcal C}$. Let us illustrate this point:

The infinitesimal transformation $\Delta t =0$ and $\Delta x= \epsilon$ is an exact symmetry (${\mathcal F}=0$) of the action constructed with the Lagrangian ${\mathcal L} =\frac12 m {\dot x}^2 e^{\lambda t}$ and gives the constant of motion ${\mathcal C}=m {\dot x} e^{\lambda t} \epsilon$. As explained above, the same constant of motion arises from the infinitesimal transformation $\Delta t = - e^{\lambda t } \epsilon$ and $\Delta x=0$, which is a symmetry (${\mathcal F} = \lambda m  x e^{\lambda t} \epsilon$) of the action constructed with the Lagrangian ${\mathcal L} = m({\dot x} \ln|{\dot x}| - \lambda x)$. Thus, the same constant of motion can result from two different symmetries.  

Third, note that if we insert~\eqref{relevant1} and~\eqref{relevant2} in~\eqref{def_C}, we get the relation as follows:
\begin{eqnarray}\label{last_relation}
{\mathcal F} =  \frac{\partial {\mathcal L}}{\partial {\dot q}^i} M^{ij} \frac{\partial {\mathcal C}}{\partial {\dot q}^j} - \frac{1}{n} {\mathcal L} \frac{\partial}{\partial {\dot q}^i} \left ( M^{ij} \frac{\partial {\mathcal C}}{\partial {\dot q}^j} \right ) - {\mathcal C}.
\end{eqnarray}
Alternatively, this relation can be obtained by substituting~\eqref{delta_q} and~\eqref{relevant1} in the constant of motion ${\mathcal C}$ expressed as~\eqref{CM1}.\footnote{In particular, for infinitesimal symmetry transformations for which $\Delta t=0$, the relation~\eqref{last_relation} acquires the simple form
$
{\mathcal C} + {\mathcal F}=  \frac{\partial {\mathcal L}}{\partial {\dot q}^i} M^{ij} \frac{\partial {\mathcal C}}{\partial {\dot q}^j}
$.}

{\bf Second relations}. From Noether's identity~\eqref{NC1}, we immediately get off-shell
\begin{eqnarray}\label{NI4}
\delta {\mathcal L}= \frac{d}{dt} \left ( {\mathcal F} - {\mathcal L} \Delta t \right ), 
\end{eqnarray}
which has a simple and deep character. This expression is also off-shell equivalent to 
\begin{eqnarray}\label{NI5}
\Delta {\mathcal L} + {\cal L} \frac{d}{dt} \Delta t
= \frac{d {\mathcal F}}{d t},
\end{eqnarray}
which, in an explicit form, reads
\begin{align}\label{alternative_identity}
\frac{\partial {\cal L}}{\partial q^k} \Delta q^k  + \frac{\partial {\cal L}}{\partial {\dot q}^k } \left [  \frac{d}{dt} \Delta q^k  - {\dot q}^k  \frac{d}{dt} \Delta t \right ] + \frac{\partial {\cal L}}{\partial t} \Delta t 
+ {\cal L} \frac{d}{dt} \Delta t
= \frac{d {\mathcal F}}{d t}.
\end{align}
Furthermore, because $\Delta q^i$ and $\Delta t$ are independent of the velocities ${\dot q}^i$, the left-hand side of~\eqref{alternative_identity} does not involve accelerations ${\ddot q}^i$ at all. The same holds for the right-hand side, because ${\mathcal F}$ is independent of ${\dot q}^i$. The reader must confront~\eqref{alternative_identity} with~\eqref{NC1} and~\eqref{NC2}. The alternative form of Noether's identity~\eqref{alternative_identity} is the departure to find all continuous symmetries (also known as variational symmetries) of the action that is constructed with a given Lagrangian ${\mathcal L}$. In other words,~\eqref{NC1},~\eqref{NC2},~\eqref{NI4},~\eqref{NI5}, and~\eqref{alternative_identity} are five equivalent ways to write Noether's identity, which is a very well-known fact. However, this is not the end of the story. Surprisingly, Noether's identity implies new relations. 

In fact, using the fact that $\Delta t$, $\Delta q^i$, and ${\mathcal F}$ are independent of ${\dot q}^i$, from~\eqref{alternative_identity} we get 
\begin{align}\label{previous_relevant5}
& \frac{\partial M_{ij}}{\partial q^k} \Delta q^k  + \frac{\partial M_{jk}}{\partial {\dot q}^i}  \left [  \frac{d}{dt} \Delta q^k - {\dot q}^k \frac{d}{dt} \Delta t \right ] + M_{jk} \left [ \frac{\partial }{\partial q^i} \Delta q^k - {\dot q}^k \frac{\partial}{\partial q^i} \Delta t  \right ]     + M_{ik} \left [ \frac{\partial }{\partial q^j} \Delta q^k - {\dot q}^k \frac{\partial}{\partial q^j} \Delta t  \right ] \nonumber\\
& + \frac{\partial M_{ij}}{\partial t} \Delta t - M_{ji} \frac{d}{d t} \Delta t =0,
\end{align}
with $M_{ij} =\frac{\partial^2 {\mathcal L}}{\partial {\dot q}^i \partial {\dot q}^j}$. Using $M_{ij}=M_{ji}$ and $\frac{\partial M_{ij}}{\partial {\dot q}^k} = \frac{\partial M_{ik}}{\partial {\dot q}^j}$, we write~\eqref{previous_relevant5} as follows:
\begin{align}\label{relevant5}
\Delta M_{ij} + M_{ik} \frac{\partial}{\partial q^j}  \Delta q^k + M_{jk} \frac{\partial}{\partial q^i}  \Delta q^k 
- M_{ik} {\dot q}^k  \frac{\partial}{\partial q^j}  \Delta t  - M_{jk} {\dot q}^k  \frac{\partial}{\partial q^i}  \Delta t 
- M_{ij} \frac{d}{dt} \Delta t =0,
\end{align}
where
\begin{equation}
\Delta M_{ij} = \frac{\partial M_{ij}}{\partial q^k} \Delta q^k  + \frac{\partial M_{ij}}{\partial {\dot q}^k} \left [ \frac{d}{dt} \Delta q^k  - {\dot q}^k  \frac{d}{dt} \Delta t \right ]  + \frac{\partial M_{ij}}{\partial t} \Delta t.
\end{equation}

Alternatively, using $\Delta q^i= \delta q^i + {\dot q}^i \Delta t$, the expression~\eqref{relevant5} can equivalently be written as
\begin{eqnarray}\label{relevant6}
\Delta M_{ij} + M_{ik} \frac{\partial}{\partial q^j}  \delta q^k + M_{jk} \frac{\partial}{\partial q^i}  \delta q^k - M_{ij} \frac{d}{dt} \Delta t =0.
\end{eqnarray}

We make three remarks as follows:

First, like~\eqref{relevant1} and~\eqref{relevant2}, relation~\eqref{relevant5} or~\eqref{relevant6} also involves $M_{ij}$, its derivatives, and the infinitesimal transformation $\Delta t$ and $\Delta q^i$. This shows that the infinitesimal symmetry $\Delta t$ and $\Delta q^i$ imposes restrictions on $M_{ij}$ and vice versa. In other words,~\eqref{relevant5} or~\eqref{relevant6} is a compatibility relation between the infinitesimal symmetry transformation $\Delta t$ and $\Delta q^i$ and the components $M_{ij}$.

Second, unlike~\eqref{relevant1} and~\eqref{relevant2}, relation~\eqref{relevant5} or~\eqref{relevant6} does not explicitly involve the constant of motion ${\mathcal C}$. 

Third, if we substitute~\eqref{delta_q} and~\eqref{relevant1} in~\eqref{relevant6}, we get a compatibility relation that involves only $M_{ij}$ and the constant of motion ${\mathcal C}$. 

In summary, relations~\eqref{delta_q},~\eqref{relevant1},~\eqref{relevant2},~\eqref{last_relation},~\eqref{previous_relevant5},~\eqref{relevant5}, and~\eqref{relevant6} are the new and main results of this section. Surprisingly, these relations have not appeared in the literature before, even though Noether's first theorem was published more than a century ago~\cite{Noether,EmmyNoether,Bessel-Hagen}. A notable feature of expressions~\eqref{delta_q},~\eqref{relevant1},~\eqref{relevant2},~\eqref{previous_relevant5},~\eqref{relevant5}, and~\eqref{relevant6} is that they involve, besides the infinitesimal symmetry transformation $\Delta t$ and $\Delta q^i$, the geometrical structure $M_{ij}$. This dependence makes it natural to combine them with Helmholtz's conditions~\cite{Helmholtz1887,Douglas_paper}, which are precisely equations for $M_{ij}$. This is the second subject of this article, which is studied below, after a brief overview of Helmholtz's conditions.

%%%%%%%%%%%%%%%%
\section{Helmholtz's conditions}\label{Helmholtz_C}
At the heart of the inverse problem of mechanics are Helmholtz's conditions~\cite{Douglas_paper,Helmholtz1887}
\begin{align}
\frac{d M_{ij}}{dt} + \frac12 M_{ik} \frac{\partial f^k }{\partial {\dot q}^j}  + \frac12 M_{jk} \frac{\partial f^k }{\partial {\dot q}^i} &=0, \label{H1}\\
M_{ik} A^k{}_j - M_{jk} A^k{}_i &= 0, \label{H2}\\
M_{ij} &= M_{ji}, \label{H3}\\
\frac{\partial M_{ij}}{\partial {\dot q}^k} &= \frac{\partial M_{ik}}{\partial {\dot q}^j}, \label{H4}\\
\det (M_{ij}) & \neq  0, \quad i,j,k, \ldots =1, \dots , n, \label{H5}
\end{align}
with
\begin{eqnarray}
A^k{}_i := \frac{d}{dt} \left ( \frac{\partial f^k}{\partial {\dot q}^i} \right ) - 2 \frac{\partial f^k}{\partial q^i} - \frac12 \frac{\partial f^k}{\partial {\dot q}^l} \frac{\partial f^l}{\partial {\dot q}^i}. 
\end{eqnarray}
These conditions are necessary~\cite{Helmholtz1887} and sufficient~\cite{Douglas_paper} for the existence of a Lagrangian ${\mathcal L}(q, {\dot q}, t)$ whose Euler-Lagrange equations
\begin{eqnarray}
\frac{\partial {\mathcal L}}{\partial q^i} - \frac{d}{dt} \frac{\partial {\mathcal L}}{\partial {\dot q}^i} &=&0, 
\end{eqnarray}
lead to the equations of motion 
\begin{eqnarray}\label{em}
{\ddot q}^i = f^i (q, {\dot q},t). 
\end{eqnarray}
Note that $M_{ij}$ is the geometric structure that underlies the system's dynamics. Moreover, note that 
\begin{eqnarray}
\frac{d M}{dt} + M \frac{\partial f^i}{\partial {\dot q}^i} =0, \quad M := \det (M_{ij}),
\end{eqnarray}
also holds. However, this relation is not an independent one, but it directly follows from Helmholtz's conditions. 

%%%%%%%%%%%%%%%%
\section{Symmetries and the inverse problem of mechanics}\label{result2}
Up to this point, the main results of the article are the new relations obtained from Noether's identity and reported in Section~\ref{results}. These relations possess conceptual and theoretical significance in their own right, regardless of any specific application. A remarkable aspect of these relations is that they contain the information about the infinitesimal symmetry transformations and also contain $M_{ij}$. As Douglas showed~\cite{Douglas_paper}, 21 years after Noether's seminal work on symmetries~\cite{Noether,EmmyNoether,Bessel-Hagen}, $M_{ij}$ is the relevant object in the inverse problem of mechanics. However, Douglas's approach to the inverse problem of mechanics does not include any information about symmetries. Therefore, it is natural to combine the new relations we reported in Section~\ref{results} with Helmholtz's conditions~\cite{Douglas_paper,Helmholtz1887} because both sets of equations contain $M_{ij}$. We use the new relations to restrict the general solution of Helmholtz's conditions~\eqref{H1}--\eqref{H5} so that the actions constructed from the resulting Lagrangians inherently exhibit the imposed symmetry. 

This leads us to essentially two new approaches or methods as follows:

{\bf Method 1}. The compatibility relation linking the infinitesimal transformations $\Delta t$ and $\Delta q^i$ with the geometrical structure $M_{ij}$, as given in~\eqref{relevant5} or~\eqref{relevant6}, is combined directly with Helmholtz's conditions~\eqref{H1}--\eqref{H5}. This is possible due to the fact both, Helmholtz's conditions and the compatibility relations, involve $M_{ij}$. In this way, Helmholtz's conditions are supplemented from the outset with the information about the symmetry transformation encoded in the relations~\eqref{relevant5} or~\eqref{relevant6}, ensuring this way that any resulting Lagrangian (and hence the corresponding action) comes endowed with the imposed symmetry. Clearly, Method 1 is more powerful than the usual approach to the inverse problem of mechanics because Douglas' approach~\cite{Douglas_paper} just produces Lagrangians. It does not produce Lagrangians with specific symmetries from the outset. Method 1 does it.

{\bf Method 2}. The relation linking the variation of the configuration variable to both the geometrical structure $M^{ij}$ and the constant of motion ${\mathcal C}$, as given in~\eqref{delta_q}, is combined with Helmholtz's conditions~\eqref{H1}--\eqref{H5}. Unlike Method 1, this approach also incorporates the constant of motion ${\mathcal C}$, which must be obtained via Noether's first theorem applied to the action constructed from the Lagrangian that encodes the imposed symmetry $\Delta t$ and $\Delta q^i$. As a result, this new method is more restrictive than Method 1. It not only looks for a Lagrangian (or a family of Lagrangians) exhibiting the imposed symmetry $\Delta t$ and $\Delta q^i$, but also ensures that the resulting action yields the constant of motion ${\mathcal C}$ through the application of Noether's first theorem. 

In the next section, we apply these two new methods.

%%%%%%%%%%%%%%%%
\section{Examples}

{\bf Example 1}. As an illustration of method 1, let us consider a single particle under the action of a frictional force proportional to the velocity (also known as damped ``free'' particle), whose equation of motion is ${\ddot x} =  -\lambda  {\dot x}$, where $\lambda$ is a positive constant. For this system, Helmholtz's condition~\eqref{H1} reduces to
\begin{eqnarray}
\frac{d M_{11}}{dt} - \lambda M_{11} =0,
\end{eqnarray}
which has the general solution $M_{11}= G (\xi_1,\xi_2)/{\dot x}$, where $G$ is an arbitrary (differentiable) function of the quantities $\xi_1={\dot x} + \lambda x$ and $\xi_2={\dot x} e^{\lambda t}$ (\cite{Gerardo_book}). Using the solution of $M_{11}$, the compatibility relation between $\Delta x$ and $\Delta t$ with $M_{11}$~\eqref{previous_relevant5} becomes
\begin{align}\label{Ex1_condition}
 &   \left[ -\frac{G}{{\dot x}^2}  + \frac{1}{{\dot x}} \left( \frac{\partial G}{ \partial \xi_1} + e^{\lambda t} \frac{\partial G}{ \partial \xi_2} \right) \right] \left(  \frac{d}{dt} \Delta x - {\dot x} \frac{d}{dt} \Delta t \right) + \frac{2G}{{\dot x}}  \left(  \frac{\partial}{\partial x} \Delta x - {\dot x} \frac{\partial}{\partial x} \Delta t \right)  + \lambda \left( \frac{\partial G}{ \partial \xi_1} \frac{ \Delta x}{\dot x} + e^{\lambda t} \frac{\partial G}{ \partial \xi_2} \Delta t \right) 
 \nonumber\\
&-\frac{G}{{\dot x}}  \frac{d}{dt} \Delta t = 0.
\end{align}
This shows that the infinitesimal symmetry $\Delta t$ and $\Delta x$ imposes restrictions on the general function $G$. In particular, for the infinitesimal symmetry transformation $\Delta t=0$ and $\Delta x=\epsilon$, the restriction~\eqref{Ex1_condition} leads to $\frac{\partial G}{ \partial \xi_1}=0$, implying $G=G(\xi_2)$ only. Consequently, $M_{11}= G(\xi_2)/{\dot x}$, with $G$ an arbitrary (differentiable) function of $\xi_2$, is compatible with this infinitesimal transformation. Two convenient choices illustrate the result. Choosing $G(\xi_2)=1$ gives $M_{11}=1/{\dot x}$, which leads to the Lagrangian ${\mathcal L} = m({\dot x} \ln|{\dot x}| -  \lambda x)$ whereas $G(\xi_2)=\xi_2$ yields $M_{11}=e^{\lambda t}$, which leads to the well-known Lagrangian ${\mathcal L} =\frac12 m {\dot x}^2 e^{\lambda t}$~\cite{Bateman}. In both cases, the actions constructed from the corresponding Lagrangians are invariant under the infinitesimal transformation $\Delta t =0$ and $\Delta x= \epsilon$, as expected. Here, the constant $m$ has been introduced to preserve homogeneity with the Lagrangians presented in Section~\ref{results}. 

If we instead consider the infinitesimal symmetry transformation $\Delta t=\epsilon$ and $\Delta x=0$, the compatibility relation~\eqref{Ex1_condition} implies $\frac{\partial G}{ \partial \xi_2}=0$, so that $G=G(\xi_1)$ only and hence $M_{11}= G(\xi_1)/{\dot x}$. In particular, the choice $G(\xi_1)=1$ recovers $M_{11}=1/{\dot x}$, which leads to the Lagrangian ${\mathcal L} = m({\dot x} \ln|{\dot x}| -  \lambda x)$ whereas the particular choice $G(\xi_1)=\xi_1$ yields $M_{11}= 1+\lambda x/ {\dot x}$, which leads to the Lagrangian ${\mathcal L} =\frac12 m {\dot x}^2+m \lambda x {\dot x} (\ln|{\dot x}|-1)-\frac12 m \lambda^2 x^2$. In both cases, the actions constructed from the corresponding Lagrangians are invariant under the infinitesimal transformation $\Delta t=\epsilon$ and $\Delta x=0$, as desired.  

Let us now consider the infinitesimal transformation $\Delta t=0$ and $\Delta x=e^{-\lambda t} \epsilon$. The compatibility relation~\eqref{Ex1_condition} implies $G-\xi_2 \frac{\partial G}{ \partial \xi_2}=0$, whose general solution is $G=\xi_2 \Phi(\xi_1)$, where $\Phi$ is an arbitrary (differentiable) function of $\xi_1$. Hence $M_{11}= \xi_2 \Phi(\xi_1)/\dot{x}$. If we choose $\Phi(\xi_1)=1$, we obtain $M_{11}=e^{\lambda t}$, which leads to the well-known Lagrangian ${\mathcal L} =\frac12 m {\dot x}^2 e^{\lambda t}$~\cite{Bateman}. As desired, the action constructed from this Lagrangian is invariant under the infinitesimal transformation $\Delta t=0$ and $\Delta x=e^{-\lambda t} \epsilon$.

{\bf Example 2}. As an illustration of method 1, let us consider a two-dimensional harmonic oscillator with the same constant angular frequency $\omega$, whose equations of motion are as follows:
\begin{eqnarray}\label{eom_rho}
{\ddot q}^1 = - \omega^2 q^1, \quad {\ddot q}^2 =- \omega^2 q^2.
\end{eqnarray}
We would like to construct Lagrangians that lead to the previous equations of motion such that their corresponding actions be invariant under the infinitesimal `small' transformation:
\begin{eqnarray}\label{symmetry_rho}
\Delta t =0, \quad \Delta q^1 = q^2 \epsilon, \quad \Delta q^2 = - q^1 \epsilon,
\end{eqnarray}
corresponding to an infinitesimal rotation. 

According to method 1, the components $M_{ij}$ must satisfy~\eqref{previous_relevant5}, which acquires the explicit form
\begin{align}\label{system_symm_rho}
q^2 \frac{\partial M_{11}}{\partial q^1} - q^1 \frac{\partial M_{11}}{\partial q^2} + {\dot q}^2 \frac{\partial M_{11}}{\partial {\dot q}^1} - {\dot q}^1 \frac{\partial M_{12}}{\partial {\dot q}^1} - 2 M_{12}=0, \nonumber\\
q^2 \frac{\partial M_{22}}{\partial q^1} - q^1 \frac{\partial M_{22}}{\partial q^2} + {\dot q}^2 \frac{\partial M_{21}}{\partial {\dot q}^1} - {\dot q}^1 \frac{\partial M_{22}}{\partial {\dot q}^1} + 2 M_{21}=0, \nonumber\\
q^2 \frac{\partial M_{12}}{\partial q^1} - q^1 \frac{\partial M_{12}}{\partial q^2} + {\dot q}^2 \frac{\partial M_{21}}{\partial {\dot q}^1} - {\dot q}^1 \frac{\partial M_{22}}{\partial {\dot q}^1} + M_{11} - M_{22} =0, \nonumber\\
q^2 \frac{\partial M_{21}}{\partial q^1} - q^1 \frac{\partial M_{21}}{\partial q^2} + {\dot q}^2 \frac{\partial M_{11}}{\partial {\dot q}^2} - {\dot q}^1 \frac{\partial M_{12}}{\partial {\dot q}^2} + M_{11} - M_{22} =0.
\end{align}
Note that the fourth equation can be obtained from the third using $M_{12}=M_{21}$, $\frac{\partial M_{21}}{\partial {\dot q}^1} = \frac{\partial M_{11}}{\partial {\dot q}^2}$, and $\frac{\partial M_{22}}{\partial {\dot q}^1} = \frac{\partial M_{12}}{\partial {\dot q}^2}$ in full agreement with~\eqref{relevant5}.

The system of equations~\eqref{system_symm_rho}, together with Helmholtz's conditions~\eqref{H1}-\eqref{H5}, has a unique solution, given by
\begin{eqnarray}\label{solution_rho}
M_{11} = c, \quad M_{12}=0, \quad M_{22}=c,  
\end{eqnarray}
where $c$ is a nonvanishing real constant. 

Using~\eqref{solution_rho}, Douglas theorem~\cite{Douglas_paper} allows us to get the only Lagrangian:
\begin{eqnarray}
{\mathcal L} = \frac{c}{2}  \left [ ({\dot q}^1)^2 + ({\dot q}^2)^2 - \omega^2(q^1)^2 - \omega^2 ({q^2})^2 \right ],
\end{eqnarray}
up to the addition of a term of the form $df(q^1,q^2,t)/dt$, that leads to the equations of motion~\eqref{eom_rho} and whose action has the desired symmetry~\eqref{symmetry_rho}.

{\bf Example 3}. As method 2 demands, relation~\eqref{delta_q} must be combined with Helmholtz's conditions~\eqref{H1}-\eqref{H5} to determine whether, given an infinitesimal transformation $\Delta q^i$ and $\Delta t$ and a constant of motion ${\mathcal C}$, there exists a Lagrangian ${\mathcal L}$ whose action is both invariant under that transformation and yields ${\mathcal C}$. 

This issue can be immediately illustrated in systems that have a single configuration variable $q$. In this case, from~\eqref{delta_q}, we get:
\begin{eqnarray}
M_{11} = \left ( \frac{1}{\delta q} \right ) \frac{\partial {\mathcal C}}{\partial {\dot q}} = \left ( \frac{1}{\Delta q - {\dot q} \Delta t} \right ) \frac{\partial {\mathcal C}}{\partial {\dot q}}, 
\end{eqnarray}
so that $M_{11}$ is entirely determined by $\Delta t$, $\Delta q^i$, and ${\mathcal C}$. Next, we must verify whether this expression satisfies Helmholtz's conditions. If it does, there exists a Lagrangian (and an action) with the desired properties; if not, such a Lagrangian does not exist.

In systems with more than one configuration variable $q^i$, we must verify the compatibility of~\eqref{delta_q} with Helmholtz's conditions. As in the one-dimensional case: if there is compatibility, then there exists a Lagrangian (and an action) with the desired properties. If there is no such compatibility, then such a Lagrangian does not exist.

%%%%%%%%%%%%%%%
\section{Concluding remarks}\label{CR}

In this article, we have shown that the inverse problem of mechanics~\cite{Douglas_paper} can be refined by incorporating symmetry requirements from the outset. By deriving novel relations directly from Noether’s first theorem~\cite{Noether,EmmyNoether,Bessel-Hagen}, we have linked the Hessian matrix, infinitesimal symmetry transformations, and their associated constants of motion into a unified framework. These new relations (reported in Section~\ref{results}), combined with Helmholtz’s conditions~\cite{Helmholtz1887,Douglas_paper}, enable the construction of Lagrangians that not only reproduce the equations of motion but also inherently embed the desired invariance. 

To broaden the scope, we highlight natural generalizations and applications:

For regular Lagrangians, and considering rigid infinitesimal transformations involving velocities~\cite{olver_book,Harvey} $\Delta t = \Delta (q,{\dot q}, t, \epsilon)$ and $\Delta q^i = \Delta q^i (q, {\dot q}, t, \epsilon)$, with linear dependence on the constant parameters $\epsilon^a$, the expressions~\eqref{NC1},~\eqref{NC2}, and~\eqref{def_C} still hold. However, in this case ${\mathcal F} = {\mathcal F} (q, {\dot q}, t, \epsilon)$ and, instead of~\eqref{delta_q}, we have
\begin{equation}\label{delta_q2}
\delta q^i = M^{ij} \left [ \frac{\partial}{\partial {\dot q}^j }  \left ({\mathcal C} + {\mathcal F } \right ) - \frac{\partial {\mathcal L}}{\partial {\dot q}^k} \frac{\partial \Delta q^k}{\partial {\dot q}^j} - \left (  {\mathcal L} - \frac{\partial {\mathcal L}}{\partial {\dot q}^k} {\dot q}^k \right ) \frac{\partial \Delta t}{\partial {\dot q}^j } \right ].
\end{equation}
Similarly, instead of~\eqref{relevant5} or~\eqref{relevant6}, we get a different expression that requires~\eqref{em} to eliminate the accelerations ${\ddot q}^i$. We will analyze this case in detail in a subsequent article.  

For singular Lagrangians, where $\det (M_{ij})=0$, the expression~\eqref{alternative_identity} also holds. Consequently, the compatibility relations~\eqref{relevant5} and~\eqref{relevant6} remain valid in this case. Thus, equation~\eqref{alternative_identity} can be used to find the variational symmetries of a given singular Lagrangian ${\mathcal L}$. Furthermore, expressions~\eqref{relevant5} and~\eqref{relevant6} can also be used to find a (singular) Lagrangian for given rigid symmetries $\Delta t$ and $\Delta q^i$. A complete analysis of singular Lagrangians, including gauge symmetries, will be presented in a subsequent article.

For one-dimensional systems, the new relations obtained from Noether's identity and reported in Section~\ref{results} can also be combined with the Jacobi last multiplier approach to the inverse problem of mechanics, because in such a case $M_{ij}$ becomes essentially the Jacobi last multiplier~\cite{Nucci_2008,Nucci_2009}. For systems with more than one variable, the relation between the results reported in Section~\ref{results} and the Jacobi last multiplier approach is unclear to us.

Finally, the extension of the current results to field theory is possible by addressing the usual technical subtleties involved (see~\cite{Santilli_book,Farias} for the analog of Helmholtz's conditions in field theory. Additionally, see~\cite{Scholle} for an approach to constructing Lagrangians in field theory that is not based on Helmholtz's conditions). In particular, we recall that a symmetric and gauge-invariant energy-momentum tensor for Maxwell and Yang-Mills fields can be obtained using only Noether's first theorem, without invoking Belinfante's method~\cite{Montesinos_Flores,Blaschke,Baker}.

\section*{Acknowledgments}
We thank Jos\'e David Vergara and Ira\'{\i}s Rubalcava-Garc\'{\i}a for their valuable comments. We thank the reviewers for drawing our attention to Refs.~\cite{Bateman,Nucci_2008,Nucci_2009,Scholle}. Diego Gonzalez acknowledges the financial support of Instituto Polit\'ecnico Nacional (Grant SIP-20253696) and the postdoctoral fellowship from Consejo Nacional de Humanidades, Ciencia y Tecnología (CONAHCyT), México.

\bibliographystyle{apsrev4-1}
	
	\bibliography{References}
	
\end{document}